\documentclass[a4paper,fleqn,usenatbib]{mnras}

\usepackage[T1]{fontenc}
\usepackage{ae,aecompl}

\usepackage{graphicx}	
\usepackage{amsmath}	
\usepackage{amssymb}	

\usepackage{ulem}
\usepackage{multicol}

\newcommand{\be}{\begin{equation}}
\newcommand{\ee}{\end{equation}}

\newcommand{\LA}{\left\langle}
\newcommand{\RA}{\right\rangle}

\title[Ejection of supermassive black holes and implications for mergers]{Ejection of supermassive black holes and implications for merger rates in fuzzy dark matter haloes}

\author[Amr A. El-Zant, Zacharias Roupas, Joseph Silk]{ 
Amr A. El-Zant$^{1}$\thanks{E-mail: amr.elzant@bue.edu.eg}, Zacharias Roupas$^{1}$, Joseph Silk$^{2,3,4}$
\\
$^{1}$ Centre for Theoretical Physics, The British University in Egypt, Sherouk City 11837, Cairo, Egypt\\
$^{2}$ Institut d’Astrophysique de Paris, UMR 7095
CNRS, Sorbonne University, 98bis Boulevard Arago, 75014 Paris, France \\
$^{3}$ The Johns Hopkins University, Department Department of Physics \& Astronomy, 3400 N. Charles Street, Baltimore, MD 21218, USA\\
$^{5}$ Beecroft Institute for Cosmology and Particle Astrophysics, University of Oxford, Keble Road, Oxford OX1 3RH, UK
}

\date{Accepted XXX. Received YYY; in original form ZZZ}

\begin{document}
\label{firstpage}
\pagerange{\pageref{firstpage}--\pageref{lastpage}}

\maketitle
	
\begin{abstract}
	Fuzzy dark matter (FDM) consisting of ultra-light axions has been  invoked to  alleviate galactic-scale problems in the cold dark matter scenario. FDM fluctuations, created via the superposition of waves, can  impact the motion of a central supermassive black hole (SMBH) immersed in an FDM halo. The SMBH will undergo a random walk, induced by  FDM fluctuations, that can result in its ejection from the central region. This effect is strongest in dwarf galaxies, accounting for wandering SMBHs and the low detection rate of AGN in dwarf spheroidal galaxies. In addition, a lower bound on the allowed axion masses is inferred both for  Sagitarius $A^*$ and heavier SMBH; to avoid ejection from the galactic centres, axion masses of the order of $10^{-22}{\rm eV}$ or lighter are excluded. Stronger limits are inferred for merging galaxies.
	We find that the event rate of SMBH mergers in FDM haloes and the associated SMBH growth rates can be  reduced by at least an order of magnitude.
\end{abstract}






\section{Introduction}

The predictions of the cold dark matter (CDM) scenario 
of structure formation have been well developed and are 
highly successful on large scales (\citealp{White_F2012}).  
CDM  normally involves weakly interacting massive particles (WIMPs)
that, assuming early thermal equilibrium,  can be produced with the right
abundance when the associated cross-section  is 
of the order expected for standard weak interactions, 

Yet extensive direct detection experiments and collider searches 
have constrained the expected  parameter space for such particles 
\citep{Wimp2018, DM_Coll2018, Waning_Wimp2018}. 
From the astrophysical point of view, WIMP-based self-gravitating 
structures (haloes) suffer from several 
'small scale' issues; 
such as the so called cusp-core problem, the overestimate of subhalo numbers, the observed diversity of rotation curves and star content, 
and the mismatch of their dynamics, in the context of the 'too big to fail' problem
(e.g., \citealp{delPopolo2017}; \citealp{Bullock_B2017} for reviews).  

We note in passing that recent studies address improved satellite modellimg that  ameliorates many of these issues, including the core-cusp issue via non-sphericity of the stellar velocity distribution \cite{hayashi2020} and the detectability of MWG satellites \cite{nadler2020}.
Other proposed solutions include those invoking baryonic physics,
ranging from inclusion of baryon-contraction-induced diversity \cite{lazar2020},
through dynamical friction-mediated coupling 
with baryonic clumps
\citep{Zant2001, Zant2004, Tonini2006, RomanoDiaz2008, Goerdt2010, Cole2011, delPopolo2014, Nipoti2015}, 
or through dynamical feedback  driven by starbursts or active galactic nuclei~\citep{Read2005, Mashchenko2006, Mashchenko2008, Peirani2008, Pontzen2012, Governato2012, Zolotov2012, Martizzi2013, Teyssier2013, Pontzen2014, Madau2014, Ogiya2014, EZFC, Silk2017, Freundlich2019}.
Alternatively modifications to the particle physics model of the dark matter
have been proposed. Such proposals include 'preheated'  
warm dark matter \citep[e.g., ][]{Colin2000, Bode2001, Schneider2012, Maccio2012b, Shao2013, Lovell2014, El-Zant2015} and
self-interacting dark matter, 
whereby energy flows into the central cores of haloes 
through conduction~\citep[e.g.,][]{Spergel2000,Burkert2000,Kochanek2000,Miralda2002, Peter2013, Zavala2013, Elbert2015}.
Ultra-light axions,  with boson mass $\sim 10^{-22} {\rm eV}$, 
have also been considered as dark matter candidates in connection with these same 
small (sub)galactic scale problems
(e.g., \citealp{Peebles2000, Goodman2000, Hu2000,  Schive2014, Marsh2014,  Hui_etal2017,
BaldiLyman, Mocz2019}; see \citealp{NiemRev19} for recent review).
Here the zero-point momentum associated with a long de Broglie wavelength
corresponding to the small mass comes along with 'fuzziness' in particle positions. 
This in turn leads to a hotter halo
core with non-diverging central density and a cut-off in halo mass. 
Such axion fields can also be relevant for inflationary scenarios 
or late dark energy models. The non-thermal production 
implies that the axions are present with the required abundance for dark matter; they behave as cold dark matter
on larger scales despite the tiny masses~\citep{Marsh2016, Marsh2017}.

The scenario is appealing in principle, despite
problems involving  the apparent  mismatch  between the 
scaling of core radii with mass inferred from simulations
and the observed scalings 
(\citealp{Hertz2018,BarnBH19,Robles2019, MohSperg19, Burkert20}).
Constraints also come from Lyman-$\alpha$ and 21 cm observations  (e.g., 
\citealp{Lyman1_2017};  \citealp{21cm1_2018}; \citealp{Hui_21cm}); 
from  environments around  supermassive black holes (e.g., \citealp{Superad}; \citealp{BarnBH19}; \citealp{NusserBH19}; \citealp{EllioMocz19}) and superradiant instability \citep{Baumann_2019,Brito_2015}; and most recently from the abundances of MWG satellites \cite{nadler2020}. 
 
A large de Broglie wavelength also leads to significant fluctuations in the density 
and gravitational fields, as bound substructures are replaced by extensive interference patterns \citep[e.g., ][]{Schive2014I}. 
These  may have observable effects on the stellar dynamics 
of galaxies, and constraints on the axion masses can in principle be  inferred
(\citealp{Hui_etal2017}; \citealp{BOFT}; \citealp{AmLoeb2018}; \citealp{Marsh2018}~\citealp{Church_2018, EZFCH}).  
A balance must therefore be struck for the FDM 
fluctuations to be large enough to solve the small-scale problems of CDM, while 
not affecting the dynamics in a manner that conflicts with observations. 
If suppression  of fluctuations is required to the extent that  implied axion masses  are too large, 
then FDM may not be useful for solving such problems. 

Another  potentially observable consequence of FDM haloes concerns the 
effect of the fluctuations on supermassive  black holes  (SMBH) inhabiting the centres of  galaxies.
In particular, the fluctuations are expected to lead 
to Brownian motion of such black holes, limited by accompanying dissipation  in the form of dynamical friction. 
Here, we make use of this fact in order to 
study the motion of a central black hole immersed in the heat bath 
of a fuzzy axion halo. In doing this, we employ the results of the model 
derived in detail recently by~\cite{EZFCH} to  describe the FDM fluctuations, assuming
a balance between fluctuation and dissipation as a result of energy equipartition. 
Naturally, in this context, we 
will be especially interested in whether the fluctuations 
can have observable effects 
on the displacement of the SMBH from the centre, and the consequences one can draw  concerning the ultra-light axion-based model of structure formation. 
Of particular interest will be constraints on the axion mass 
inferred from the empirical finding of 
well-centered active galactic nuclei (AGN), especially in massive galaxies, 
and off-centred AGN, sometimes out to  kpc scales in less massive galaxies 
(\citealp{Menez_off_cent14, Menez_off_cent16, Reine_wandering_20, Shen_2019}). 

In the following section, we present the FDM density profile we use in our analysis and our pre-assumed SMBH-halo scaling relations. In section \ref{sec:expulsion}, we calculate the SMBH velocity dispersion, the relaxation time-scale and the displacements of the random walks induced by the FDM fluctuations. We also estimate the event reduction rate in SMBH mergers. In the two final sections, 
we discuss our conclusions.

\section{Fuzzy Dark Matter Halo Profiles}
\label{sec:FDM_profile}

\subsection{SMBH-halo scaling relations}
\label{sec:scaling}
Well-known  relations connect the masses of SMBHs at the 
centres of galaxies to central velocity dispersions and to masses of the bulges
of galaxies. The BH masses are further connected to those of the 
host haloes both observationally and through simulations 
(e.g.,~\citealp{Ferrarel02, Simardrel09, ComBH19, ilusstruBHrel18}). 
With the results of such investigations in mind,
we use the following formula to relate the halo mass to the SMBH mass
\begin{equation}
M_B = 10^7 M_\odot \left( \frac{M_H}{10^{12} M_\odot} \right)^{3/2}.     
\label{eq:MBMH}
\end{equation}
where, strictly speaking, $M_H$ here  corresponds to the virial mass 
of a cosmological halo. In addition, dissipationless
structure formation simulations suggest that there is a strong corelation between the 
typical velocities of dark matter particles and the mass of the 
halo, and is of the form $M_H \sim \sigma^3$ (e.g., \citealp{Evrardhaloscal08, Klypin2011}).
For example,~\cite{Evrardhaloscal08}, 
find  for the one-dimensional velocity dispersion  
of the halo (averaged over all particles)
$\sigma_H = 108.3 \left(\frac{h (z)  M_H} {10^{12} M_\odot}\right)^{0.334}$, where $h$ is the scaled Hubble parameter;  
$h (z = 0) \approx 0.7$. In this context, we adopt
\begin{equation}
M_H  = 10^{12} M_\odot \left(\frac{\sigma_H} {100 {\rm km/s}}\right)^{3}.
\label{eq:HMSIG}
\end{equation}

\subsection{Soliton core}
\label{sec:corescal}

\begin{figure}
	\includegraphics[width=0.49 \textwidth]{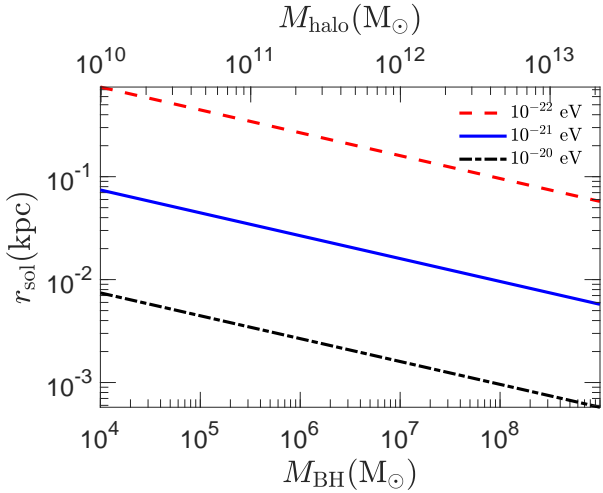}
	\caption{ The soliton core radius with respect to the SMBH mass and the FDM halo mass for three different axion masses. Our analysis is valid in the region $r\gtrsim r_{\rm sol}$.
	}
	\label{fig:r_sol}
\end{figure}

Cosmological simulations of FDM suggest that FDM haloes contain a solitonic central core whose density profile is empirically fit by the formula \citep{Schive2014I}
\begin{equation}
	\rho_\text{core} = \frac{\rho_{\rm sol}}{\left[1 +0.091~\left(r/r_{\rm sol}\right)^2\right]^8},
	\label{eq:rho_core}
\end{equation}
where the soliton core radius can be expressed as  
\begin{equation}
\label{eq:r_sol}
	r_{\rm sol} = 0.16 {\rm kpc} \left( \frac{m}{10^{-22} {\rm eV}}  \right)^{-1} \left(\frac{M_H}{10^{12} {\rm M_\odot}} \right)^{-1/3},
\end{equation}
and $M_H$ is the halo, virial, mass. The characteristic soliton density is 
\begin{equation}
	\rho_{\rm sol} =  2.94 \times 10^{10} \left( \frac{m}{10^{-22} {\rm eV}}  \right)^{-2}  
	\left(\frac{r_{\rm sol}}{0.16 {\rm kpc}}\right)^{-4} \frac{\rm M_{\odot}}{\rm kpc^3}
	\label{eq:rhosol}
\end{equation}

We will consider that a SMBH, associated with the FDM halo, is on an 
orbit that initially takes it well outside the soliton core, where the calculations of~\cite{EZFCH}
are strictly valid. Therefore, our analysis applies for $r \gtrsim r_{\rm sol}$, as in Figure \ref{fig:r_sol}. 
The SMBH may,  in this context, be assumed to have been perturbed 
from the centre at some stage during the hierarchical galaxy formation scenario 
(e.g, during a major merger). 
In the case of a CDM halo  it then spirals in by dynamical friction
towards the very centre. In contrast, in the FDM scenario we consider here,  
the interplay and competition 
between fluctuations and dissipation results in the SMBH  distance from  the centre statistically satisfying  energy equipartition~(\ref{eq:equipar}).
One main purpose here will be to investigate the stalling of SMBH mergers 
that accompany galactic mergers. 
In the context of this scenario, it will not be of crucial importance to model
the density distribution precisely inside $r_{\rm sol}$. We assume 
that most FDM bound to the merging SMBH has been stripped and is dynamically 
unimportant, though we briefly discuss (in Section~\ref{sec:inter}) 
the role of  nuclear stellar clusters, surrounding SMBHs, 
in accellerating their final merger. 

\subsection{FDM halo}
\label{sec:halo}

The profile given by~(\ref{eq:rho_core})  is not accurate outside the soliton core.  It has been suggested that
there the density can be described instead by the NFW profile, which joins smoothly with 
the above form at about three soliton radii $r_{\rm sol}$ (e.g, \citealp{Mocz2019}).  
According to recent numerical simulations of ~\cite{Veltmaat_2018}, however,  
6

6

in this region the velocities are well fitted by a Maxwellian distribution, suggesting that the density profile is nearly isothermal. 
Indeed, just outside the soliton core, the density profiles  seems flatter than NFW.
In this context, and in line with our assumption of Maxwellian distribution for velocity, we consider a profile that approximately fits a  
cored isothermal sphere, namely 
\begin{equation}
	\rho_\text{halo} = \frac{3 \sigma_H^2}{2 \pi G r_c^2}g(x) \,,
	\text{where}\;
g(x) = \frac{1 + \frac{1}{3}x^2}{\left(1 + x^2\right)^2}  
\,,
\;
x = \frac{r}{r_c}.
	\label{eq:rho_halo}
\end{equation}

In Appendix~\ref{app:corad},  we show that matching the core with the halo profiles at $3r_{\rm sol}$ gives
\begin{equation}
r_c = 2~{\rm kpc} \left(\frac{m_{\rm axion}}{10^{-22}{\rm eV}}\right)^{-1} \left(\frac{M_{\rm B}}{10^7M_\odot}\right)^{-2/9},
\label{eq:r_c}
\end{equation}
where the scaling relations (\ref{eq:MBMH}), (\ref{eq:HMSIG}) are assumed to hold.

The density profile (\ref{eq:rho_halo}) generates the logarithmic potential 
\begin{equation}
	\Phi = \sigma_H^2 \ln \left(1 + x^2\right), 
	\label{eq:Phi}
\end{equation}
and tends at large radii
to that of a singular isothermal sphere with isotropic one-dimensional velocity dispersion $\sigma_H$. Nevertheless, the velocity dispersion varies as
\begin{equation}
\label{eq:sigma_halo}
	\sigma_{\rm halo} = \sigma_H \,s(x)\,,\;
	\text{where}\;
	s(x) = \left(\frac{x^2 + 2}{x^2 + 3}\right)^{1/2},
\end{equation}
as shown in Appendix~\ref{app:velprof}. For large $x$, this tends
to unity; it differs by at most $18 \%$ from this value at the very
centre, validating our Maxwellian approximation for the velocity distribution.

\section{Ejection of SMBHs}
\label{sec:expulsion}

\subsection{SMBH displacement} 
\label{sec:disp}

\begin{figure}
	\includegraphics[width=0.49 \textwidth]{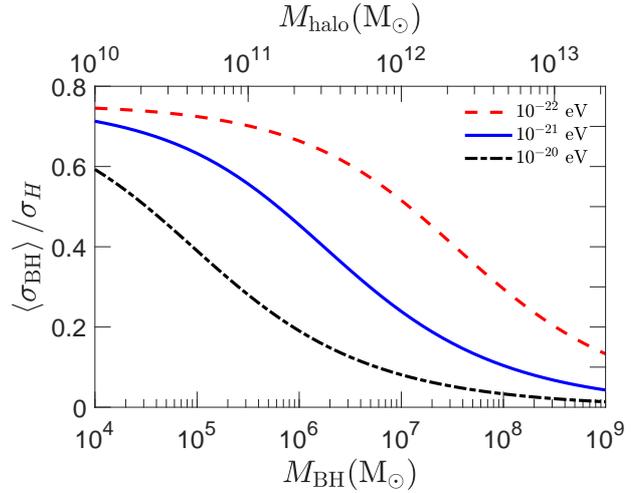}
	\caption{ The average FDM-induced SMBH velocity dispersion $\LA \sigma_{\rm BH} \RA$ scaled over the characteristic halo velocity dispersion $\sigma_H$ with respect to the SMBH and halo masses for different axion masses.
	}
	\label{fig:sigma_BHoverH}
\end{figure}

\subsubsection{General considerations and scaling of SMBH velocity dispersion}

We consider a black hole in the central region of an FDM sphere. 
The interfering  de Broglie waves of the ultra light FDM axions 
lead to density fluctuations, which are related 
though the Poisson equation to potential fluctuations. 
Their dynamical effect  on a classical particle 
can be described in a manner analogous to that of 
standard two body relaxation theory (\citealp{BOFT,EZFCH}). 
In this context, for a Maxwellian FDM velocity distribution, the FDM fluctuations 
are generated by granules of characteristic effective mass
\begin{equation}
m_{\rm eff} = \frac{ \pi^{3/2} \hbar^3}{m_{\rm ax}{}^3 \sigma^3}  \rho,
\label{eq:meff}
\end{equation}
where $m_{\rm ax}$ is the axion mass, $\rho$ is the average background density of FDM axions and
$\sigma$ their one-dimensional velcoity dispersion. 
Adopting $\rho = \rho_{\rm halo}$ given in (\ref{eq:rho_halo}) and $\sigma = \sigma_{\rm halo}$ given in (\ref{eq:sigma_halo}) we get
\begin{equation}
\label{eq:m_efff_BH}
m_{\rm eff} = 1.1\cdot 10^7 M_\odot \left(\frac{m_{\rm ax}}{10^{-22}{\rm eV}}\right)^{-1}
\left(\frac{M_B}{10^7M_\odot}\right)^{2/9} \frac{g(x)}{s(x)^3}.
\end{equation}

As detailed in Appendix~\ref{app: effective}, we 
stress that this is an effective description; the 'granules' thus described are not 
long-lived classical particles. They arise instead from a formal equivalence with fluctuations 
giving rise to standard two body relaxation. The  
FDM power spectrum of fluctuations that gives rise to this description fully matches
that inferred from numerical simulations. The effective 'size' (and hence mass) 
of the quasiparticles in fact corresponds 
to a cutoff at the characteristic de Broglie wavelength in that spectrum, which has 
white noise structure on large scales.

We assume that the SMBH achieves equilibrium with the fluctuations; that is, 
there is a balance between the effects of fluctuation and dissipation, 
the latter being due to dynamical friction
(for this to be the case the relevant  relaxation time must be small enough, as 
discussed in Section~\ref{sec:RW_tau}). 
Because of equipartition, the velocity dispersion $\sigma_B$ of the SMBH 
is related to the FDM velocity dispersion as
 \begin{equation} 
 M_{B} \sigma^{2}_{B} = m_{\rm eff}   \sigma^2_{\rm eff}
 = \frac{1}{2} m_{\rm eff} \sigma_{\rm halo}^2,
\label{eq:equipar}
\end{equation}
where $\sigma^2_{\rm eff} = \sigma_{\rm halo}^2/2$ refers to the effective velocity dispersion of the 
FDM quasiparticles with mass $m_{\rm eff}$ (cf.~\citealt{EZFCH}). Substituting the effective mass from (\ref{eq:meff}) we get the SMBH velocity dispersion
 \begin{equation} 
\sigma_{B} = 73.8 ~\frac{\rm km}{\rm s} \left(\frac{m_{\rm ax}}{10^{-22}{\rm eV}}\right)^{-\frac{1}{2}}
\left(\frac{M_B}{10^7M_\odot}\right)^{-1/6} \frac{g(x)^{1/2}}{s(x)^{1/2}}, 
\label{eq:sigmaB}
\end{equation}
and 
\begin{equation}
\frac{\sigma_B}{\sigma_{\rm halo}}
	=  0.74
		\left(\frac{m_{\rm ax}}{10^{-22}{\rm eV}}\right)^{- \frac{1}{2}}
	\left(\frac{M_B}{10^7M_\odot}\right)^{-7/18} \frac{g(x)^{1/2}}{s(x)^{3/2}}.
\label{eq:sigmarat} 
\end{equation}

This scaling with mass is a result of the empirical  relation (\ref{eq:MBMH}), which 
implies a smaller SMBH
mass relative to halo mass, 
combined with the fact that the de Broglie wavelengths (and effective masses)   
are larger for smaller haloes.

\subsubsection{Expected RMS displacement from thermal average}

The FDM halo acts as a heat bath for the SMBH, which will therefore undergo Brownian motion induced by the FDM fluctuations, 
with a characteristic radius reflecting a balance between the effect of fluctuations and that of the accompanying dissipation 
from dynamical friction. It has been shown by \cite{1988JPCS...49..673V} that the one-particle probability distribution function of Brownian motion in inhomogeneous medium (presence of temperature gradient and external potential) is given at equilibrium by (see also \citealt{1981ZPhyB..41...39R,1989PhyA..154..452W,1965JChPh..43.1110N,1968Phy....39..334Z,1994PhyA..212..231P})
\begin{equation}
\label{eq:f_prop}
f(x) = \frac{\beta_B(x) \exp\left\lbrace-\int_0^x d\tilde{x}\beta_B(\tilde{x}) M_{\rm B} \Phi(\tilde{x})^\prime d\tilde{x}\right\rbrace }{\int_0^\infty \beta_B(x) \exp\left\lbrace-\int_0^x d\tilde{x}\beta_B(\tilde{x}) M_{\rm B} \Phi(\tilde{x})^\prime d\tilde{x}\right\rbrace  4\pi x^2  dx} 
\end{equation}
where a prime denotes a derivative and $\Phi(x)$ is given in (\ref{eq:Phi}). 
For the derivation and application in self-gravitating systems see \cite{2020arXiv200612755R}.
We denote
\begin{equation}
\beta_B^{-1} \equiv m_{\rm eff}\sigma_{\rm eff}^2 = M_{\rm B} \sigma_{\rm B}^2.
\end{equation}
The probability (\ref{eq:f_prop}) is the generalization of the Boltzmann propability $f_B \propto e^{-\beta m \Phi}$ in the case of varying $\beta$ and is the stationary solution of an appropriate Fokker-Planck equation given by \cite{1988JPCS...49..673V}.
We may therefore average over $x$ any physical quantity $A(x)$ referring to the SMBH as 
\begin{equation}
	\LA A(x) \RA =  \int_0^\infty A(x)  f(x) 4\pi x^2 dx .
\label{eq:average}
\end{equation}
The average SMBH velocity dispersion, scaled over the characteristic halo velocity dispersion $\sigma_H$, is plotted in Figure \ref{fig:sigma_BHoverH}. It is higher for lighter axions and lighter SMBHs.

A statistical measure of the distance at which the SMBH may be ejected from the center is the RMS displacement, which is plotted in Figure \ref{fig:r_RMS}. The displacement is numerically fit by the expression 
 \begin{equation}\label{eq:r_RMS}
 	r_{\rm RMS} \equiv 
 	\sqrt{\LA r^2\RA} = 1.84~{\rm kpc} \left(\frac{m_{\rm ax}}{10^{-22} {\rm eV}}\right)^{-\frac{3}{2}}
\left(\frac{M_B}{10^7 M_\odot}\right)^{- \frac{3}{5}}.
 \end{equation}
 Evidently, it is higher for lighter SMBHs and lighter axions.
This effect can explain observations indicating a dearth of SMBHs in the centers 
of dwarf galaxies. 
We discuss this further in Section~\ref{sec:inter}.
In addition, we infer that the displacement of the SMBH can be so large that it is completely ejected out of the bulge. Observations of SMBHs residing deep in their host bulges allow us constrain the axion mass. Requiring $r_{\rm RMS} \lesssim 0.1{\rm kpc}$ for Sagittarius $A^{*}$, with $M_B = 4.1\cdot 10^6M_\odot$, constraints the axion mass to
$
m_{\rm ax} \gtrsim 10^{-21}{\rm eV}.
$
Requiring that the RMS displacement from the centre be less than 
$1 {\rm kpc}$, on the other hand,  leads to the constraint 
$m_{\rm ax} \gtrsim 2\cdot 10^{-22}{\rm eV}$. Finally, we have $r_{\rm RMS} > 0.1{\rm kpc}$ for all SMBHs lighter than $10^9M_\odot$, and for axions lighter than $10^{-22}{\rm eV}$.

\subsubsection{Virial Theorem} 
\label{sec:virial}
We analyze further and validate the results regarding the RMS displacement~(\ref{eq:r_RMS}), calculated as the average over the random walk (\ref{eq:f_prop}) and displayed in Figure~\ref{fig:r_RMS}, by invoking the virial theorem for the SMBH. For the potential (\ref{eq:Phi}) the virial theorem translates to 
\begin{equation}
\langle \ln (1+x^2) \rangle = \left<3\frac{ \sigma_B^2}{\sigma_{\rm halo}^2}\right>.     
\label{eq:virial}
\end{equation}
Using this equation we may estimate the RMS displacement without any need to assume a particular  probability distribution in order to obtain the average. 

This may be achieved as follows. For $x \la 1$, it is   
$\LA\ln (1+x^2) \RA\approx  \langle x^2 \rangle$. In this case, 
using~(\ref{eq:sigmarat}) and~(\ref{eq:r_c_fit})
one may estimate the RMS displacement straightforwardly
\begin{equation}
\sqrt{\LA r^2 \RA } \approx  2.6~{\rm kpc} \left(\frac{m_{\rm ax}}{10^{-22} {\rm eV}}\right)^{-\frac{3}{2}}
\left(\frac{M_B}{10^7 M_\odot}\right)^{-\frac{11}{18}} \left< \frac{g(x)}{s(x)^3} \right> ^{\frac{1}{2}}.
\label{eq:estsmallx}
\end{equation}
From~(\ref{eq:rho_halo}) and~(\ref{eq:sigma_halo}),  
one deduces that $\frac{g(x)}{s(x)^3}$ is order one for $x \la 1$. Despite that the approximation is rough, the agreement with Eq. (\ref{eq:r_RMS}) is striking. The powers of masses are nearly identical, while even the scaling factor is matching very well. This self-consistency check provides further supporting evidence for the validity of our approach.

The condition $x \la 1$, combined with 
(\ref{eq:virial}) implies that
$\frac{3 \sigma_B^2}{\sigma_{\rm halo}^2} \la 1$. 
Recalling again that $\frac{g(x)}{s(x)^3}$ is order one for $x \la 1$,  
from  (\ref{eq:sigmarat}) this condition is seen to correspond 
to SMBH masses 
$\ga 10^{7} M_\odot$, when the axion mass is $10^{-22} {\rm eV}$. 
This in turn relates to halo masses  $\ga 10^{12} M_\odot$. 
That is, the condition $x \la 1$ 
applies, in this case, to Milky Way-type haloes and above. For larger 
larger axion masses, the approximation holds for
smaller halo and SMBH masses; in general, it is valid for  
$M_B \ga 10^7 (\frac{m_{\rm ax}}{10^{-22} {\rm eV}})^{-9/7} M_\odot$. 

For smaller masses 
the RMS displacements keep on increasing, 
even though $x$ can be larger than unity.   
To get an estimate of the expected excursion in this 
case, consider again equation~(\ref{eq:virial}), 
with  $x \gg 1$. 
Now $\ln (1+ x^2) \rightarrow 2 \ln x$ and 
$\langle 3 \frac{g(x)}{s(x)^3} \rangle^{1/2}\rightarrow 
\langle x^{-2} \rangle^{1/2}
\ge  \langle x^2 \rangle^{-1/2}$. In this case,  
\begin{equation}
\sqrt{\langle r^2 \rangle} \ga \frac{2.56~{\rm kpc}}{\sqrt{6 \langle \ln x \rangle}}   \left(\frac{m_{\rm ax}}{10^{-22} {\rm eV}}\right)^{-3/2}
\left(\frac{M_B}{10^7 M_\odot}\right)^{-11/18}.
\label{eq:estlargex}
\end{equation}
Except for the weak root-logarithm dependence, 
this has the same scaling 
as in (\ref{eq:estsmallx}). The slightly weaker scaling 
is again consistent with the trend in figure~\ref{fig:r_RMS}. 

Note finally that for small SMBH masses and $m_{\rm ax} < 10^{-21} {\rm eV}$ the RMS displacement may be exceptionally large, even larger than the 
expected virial radius. Such displacement is therefore unrealistic. 
This is evident from Figure~\ref{fig:r_RMS} 
and equation~(\ref{eq: virialr}). 
We will see in Section~\ref{sec:RW_tau}
below, however, that the displacements from the centre  are limited by the large relaxation time required to maintain thermal coupling between the SMBH and the FDM heat bath 
at low densities, characteristic of the outer halo.

\subsubsection{Reduction of RMS radius by baryons}
\label{sec:baryons}

The RMS radius just evaluated reflects a statistical balance between  
fluctuations and dissipation. When the FDM mass fraction is decreased, 
as a baryonic component is introduced, the effect of fluctuations is suppressed relative to that of the dissipation. This leads to a smaller 
RMS radius at equilibrium, as we now estimate, focusing 
on the case of massive galaxies where the role of 
baryons is more prominent than in dark matter dominated dwarfs. 

The  energy dissipation rate, due to a dynamical friction 
force ${\bf F}_D$
acting on a particle moving with velocity ${\bf v}$, is ${\bf F}_D. {\bf v}$. 
For a particle of mass $m$, moving through 
field particles with masses $m_a$ and (strictly speaking homogeneous)  density $\rho$, $F_D/m$ scales  
with these variables as $(m_a + m) \rho$ 
(e.g.~\citealp{BT}). 
Therefore, if an  FDM halo hosts a system of stars 
of mass $m_*$, distributed with density 
$\rho_*$, the specific energy dissipation rate 
of an SMBH scales as  
$(m_{\rm eff} +M_B) \rho_{\rm halo} + (m_* + M_B) \rho_*$ 
(if the  FDM is treated 
as a system of particles of mass $m_{\rm eff}$). 
From equation~(\ref{eq:m_efff_BH}) one sees that $m_{\rm eff} \gg m_*$; 
and that for massive galaxies (with $M_B \ga 10^7 M_\odot$),   
$M_B \ga m_{\rm eff}$ for $m_{\rm ax} \ge 10^{-22} {\rm eV}$. 
Thus the  dissipation rate scales approximately as 
$\sim M_B (\rho_{\rm halo} + \rho_*) = M_B~\rho$. 
It is therefore largely unchanged by the 
introduction of a stellar 
component, while keeping the total density constant. 
(Dynamical friction with a gaseous, rather than stellar,  
component would also contribute in approximately 
in the same manner; \citealp{OsE_DF99}).

The rate of energy gain due to 
fluctuations is quantified through the change of velocity variance 
$\LA (\Delta v)^2 \RA$ (cf. equation~\ref{eq:disp_inc} below for the case of FDM). This does change when baryons are added. It 
scales as $m_{\rm eff} \rho_{\rm halo} + m_* \rho_*$.
And as $m_{\rm eff} \gg m_*$, the baryons don't contribute significantly to the fluctuations. 
The ratio  of the energy gain, due to fluctuation,
to energy loss, due to dissipation, then scales as
$m_{\rm eff}/M_B  \times \rho_{\rm halo}/\rho$; it decreases
as the baryonic component is introduced, because 
$\rho_{\rm halo}/\rho$ is reduced, 
The reduction is  equivalent 
to that obtained by 
keeping  $\rho_{\rm halo}/\rho = 1$ and decreasing 
$m_{\rm eff}/M_B$.

The equilibrium radius is in turn reduced. 
For, as the baryons do not participate in the 
fluctuations,  the equilibrium condition (\ref{eq:equipar}) 
is modified to $M_B \sigma_B^2 =\frac{1}{2} \frac{\rho_{\rm halo}}{\rho} 
m_{\rm eff} \sigma_{\rm halo}^2$.  
To estimate the effect for massive  galaxies  
one can use~(\ref{eq:virial}),
with  $x \la 1$ as in Section~\ref{sec:virial} above.
This gives  
$r_{\rm RMS}/r_c \approx \sqrt{3} \sigma_B/ \sigma_{\rm halo} \approx 
\sqrt{ m_{\rm eff}/ M_B \times {\rho_{\rm halo}}/{\rho}}$. 
The equilibrium radius thus decreases with FDM fraction 
as $r_{\rm RMS} \sim (\rho_{\rm halo}/\rho)^{1/2}$. 
In addition, if the total density
is kept constant, $m_{\rm eff} \propto \rho_{\rm halo}$ 
also decreases by a factor $\rho_{\rm halo}/\rho$ when baryons are added. 
The combined effect results in 
$r_{\rm RMS} \sim \rho_{\rm halo}/\rho$.

Since  (from \ref{eq:meff}) $m_{\rm eff}$ also scales as
$1/m_{\rm ax}^3$, the constraints on the axion mass 
are weakened by a factor $\sim (\rho_{\rm halo}/\rho)^{2/3}$. 
This is not very large, unless $\rho_* \gg \rho_{\rm halo}$. 
We will see in section~\ref{sec:LISA}, however, 
that out-of-equilibrium 
effects can still cause SMBHs interacting 
with baryons to inspiral towards the centre.


\subsection{Relaxation time-scale}
\label{sec:RW_tau}

\begin{figure}
	\includegraphics[width=0.49 \textwidth]{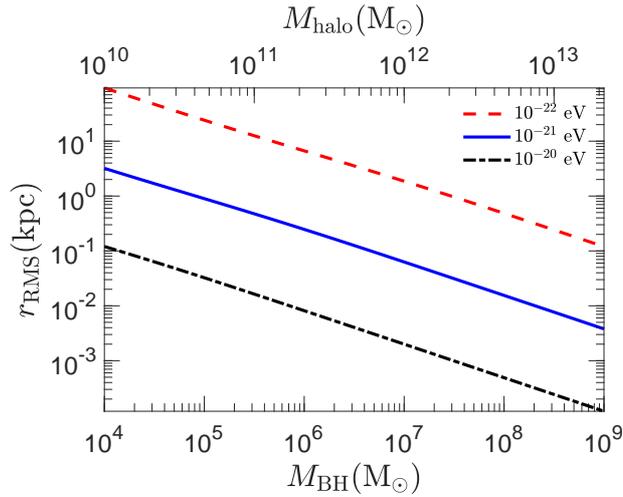}
	\caption{ The RMS displacement $\sqrt{\LA r^2\RA}$ of the SMBH random walk induced by FDM halo fluctuations.
	}
	\label{fig:r_RMS}
\end{figure}

\begin{figure}
	\includegraphics[width=0.49 \textwidth]{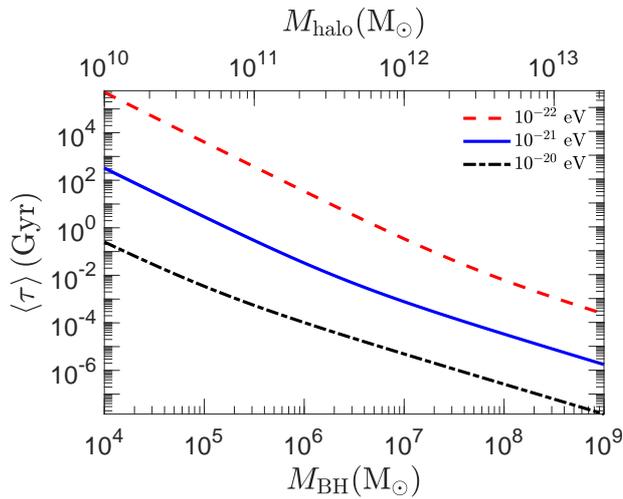}
	\caption{The average relaxation time of the SMBH-FDM halo system. 
	}
	\label{fig:tau}
\end{figure}

We calculate the relaxation time of the SMBH-FDM halo system as follows. When fluctuations are dominant over dissipation (dynamical friction) the velocity dispersion of the black hole is expected to increase with time as (see equation~(62) in \citealt{EZFCH})
\begin{equation}
\LA (\Delta v_B)^2 \RA
=  t \frac{ 8 \pi G^2 \rho_{\rm halo} m_{\rm eff} \ln \Lambda}{\sigma_{\rm halo}} \,\frac{{\rm erf}  (X)}{X}  ,
\quad 
X=\frac{v_B}{\sigma_{\rm halo}}, 
\label{eq:disp_inc}
\end{equation}
where $\ln \Lambda$ is the Coulomb logarithm (see Appendix~\ref{app:Coul}). 
The relaxation timescale of the black hole-dark matter halo system may be calculated by estimating the time it is required for $\Delta v_B \sim v_B$, when $v_B = \sigma_B\sqrt{3}$. Requiring in particular $\LA (\Delta v_B) ^2\RA = 3\sigma_B^2$  equation (\ref{eq:disp_inc}) then gives 
\begin{align}
\tau &=  \frac{3\sqrt{3}}{8 \pi G^2 \ln \Lambda }\frac{\sigma_B^3}{\rho_{\rm halo} m_{\rm eff}} \left\lbrace{\rm erf}\left(  X\right)\right\rbrace^{-1}
\nonumber\\
&= \frac{1.5 }{\ln \Lambda} ~{\rm Gyr}
\left(\frac{m_{\rm ax}}{10^{-22}{\rm eV}}\right)^{-5/2}
\left(\frac{M_B}{10^{7}M_\odot}\right)^{-29/18}
\nonumber \\
&~ \times \frac{s(x)^{3/2}}{g(x)^{1/2} {\rm erf}\left(  X\right)}
\label{eq:tau}
\end{align}
where $X\equiv	\frac{\sigma_B}{\sigma_{\rm halo}}\sqrt{3}$, 
and where we have used (\ref{eq:sigmaB}).
Using~(\ref{eq:average})  one can evaluate the average 
relaxation time $\LA\tau \RA$. 
This is shown in Figure \ref{fig:tau} where $\Lambda$ is given by Eq. (\ref{eq:Lambda}). It is higher for lighter axion masses and lighter SMBHs.
The relaxation time-scale is smaller than a Hubble time for all  $m_{\rm ax } \ge 3\cdot 10^{-21}{\rm eV}$ and $M_{\rm BH} \ge 10^4M_\odot$.
For $m_{\rm ax } = 10^{-21}{\rm eV}$ and $m_{\rm ax } = 10^{-22}{\rm eV}$ the time-scale is less than a Hubble time for $M_{\rm BH} \ge 5 \cdot 10^4M_\odot$ and $M_{\rm BH} \ge 2\cdot 10^6 M_\odot$, respectively. 
{ Self-consistency of our analysis,  namely  the use of 
equipartition conditions (\ref{eq:equipar})  and averages given by (\ref{eq:average}),
requires that this be the case.}
Therefore the Sagittarius $A^*$ and heavier SMBHs are well probed by our analysis for axion masses $\gtrsim 10^{-22}{\rm eV}$. The SMBHs of dwarfs are  probed by our analysis for axion masses as low as $\sim 10^{-21}{\rm eV}$. 

\subsection{LISA event reduction}
\label{sec:LISA}

 \begin{figure}
	\includegraphics[width=0.49 \textwidth]{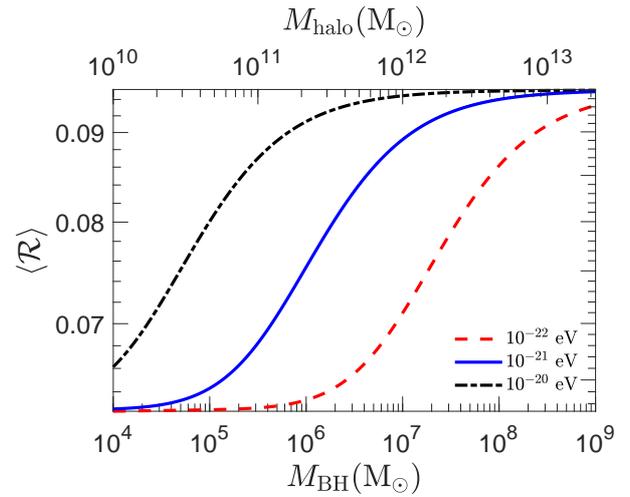}
	\caption{Estimated average event reduction rate of LISA for stellar 
	to FDM density ratio of 0.2.}
	\label{fig:Reduction}
\end{figure}

In the presence of the FDM halo, we have calculated that each SMBH will experience a random walk that will tend to keep them separated. At statistical equilibrium, the RMS displacement is given by Eq~(\ref{eq:r_RMS}) and is displayed in Figure \ref{fig:r_RMS}. The emergence of a stalling radius for SMBH mergers inside FDM haloes has been suggested already in \citep{Hui_etal2017,BOFT} in a different context. 
The stalling radius we propose here,  $r_{\rm RMS} = \sqrt{\LA r^2\RA}$, 
is statistical, as it is evaluated over the random walk Eq. (\ref{eq:f_prop}), which is a stationary solution of a Fokker-Planck equation \citep{1988JPCS...49..673V}. 
This stalling radius is not strict but may be crossed randomly if a random kick is sufficiently intense.

If baryons are added to mass distribution, 
statistical equilibrium between fluctuation 
and dissipation, is perturbed. The fluctuations are 
unchanged but the magnitude of the dissipation (dynamical friction)
is increased, and the stalling radius decreases (Section~\ref{sec:baryons}).  

Departures from equilibrium may also 
lead to the inspiral of SMBH to the centre; 
and in merging systems bring a pair of SMBHs 
within range their  combined sphere of 
influence, thus forming a binary 
(this is discussed further in Section~\ref{sec:inter}).
So the possibility of merger in an FDM scenario still exists. 
But the event rate of SMBH mergers observable by LISA should be significantly reduced. 

The FDM equipartition timescale can be a good probe of the frequency 
and strength of the random fluctuations acting on the SMBHs, thus characterizing 
the span for the restoration 
to equilibrium in the presence of perturbation.
If, for example, a large increase in dynamical friction due to coupling 
with a stellar component takes place, the restoration 
to equilibrium (with new stalling radius) will depend on the ability of the fluctuations to timely restore the system to the new equilibrium.
A situation where this could be crucial involves an SMBH entering 
the central halo region, with significant baryon density, during a 
merger. 

A first estimate of the reduction factor of the LISA event rate may thus 
involve only the ratio of the FDM equipartition timescale over the stellar dynamical friction timescale. We estimate the dynamical friction timescale as $\tau_\star \approx \sigma_B/|a_{\rm DF}|$, where $a_{\rm DF}$ denotes the dynamical friction deceleration caused by field stars. Substituting $a_{\rm DF}$ for a Maxwellian distribution as in \citep{BT} we get
	\begin{equation}\label{eq:tau_df}
	\tau_{ \star} = \frac{3\sqrt{3}}{4 \pi G^2 \ln \Lambda_\star }
	\frac{\sigma_B^3}{\rho_\star M_{\rm B}} \left\lbrace{\rm erf}\left(  X\right) -\frac{2X}{\sqrt{\pi}} e^{-X^2} \right\rbrace^{-1},
	\end{equation}
	where $\rho_\star$ is the stellar density.
	Dividing the equipartition timescale Eq. (\ref{eq:tau}) over Eq. (\ref{eq:tau_df}), we get 
	\begin{equation}
	\frac{\tau }{\tau_{\star}} = 
	\frac{1}{4}\frac{\ln \Lambda_\star}{\ln\Lambda} \frac{\sigma_{\rm halo}^2}{\sigma_{\rm B}^2} \frac{\rho_\star}{\rho_{\rm halo}}
		\left(1 - \frac{2X e^{-X^2}}{ {\rm erf}(X) \sqrt{\pi} }\right)
	\end{equation}
	We estimate the event reduction factor with the expression
\begin{equation}
	\mathcal{R} \equiv 1-e^{-\frac{\tau }{\tau_{\star}}},
\end{equation}
which is finite for all $\tau$, and converges to $\mathcal{R}\rightarrow \tau/\tau_{\star}$ for $\tau\ll \tau_{\star}$ and to $\mathcal{R}\rightarrow 1$ for $\tau \gg \tau_{\star}$. An average over the unperturbed equilibrium is implied.  
		
	In Figure \ref{fig:Reduction} we plot the average event reduction factor $\LA\mathcal{R}\RA$ with respect to SMBH mass for different axion masses, assuming $\rho_\star/\rho_{\rm halo} = 0.2$ and $\ln \Lambda_\star \approx \ln \Lambda$. It is evident that for any axion mass and any SMBH mass. The  event rate, thus deduced, is reduced by at least an order of magnitude, and is smaller for smaller systems.

 \section{Discussion}
 \label{sec:inter}

\begin{figure}
	\includegraphics[width=0.49 \textwidth]{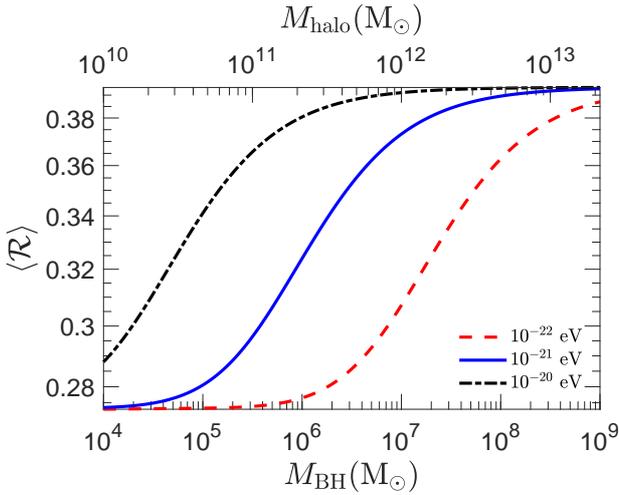}
	\caption{Same as in Figure~\ref{fig:Reduction}, but for 
	stellar to FDM density ratio of 1}
	\label{fig:Reduction1}
\end{figure}

SMBHs can be expelled from the centre of an FDM halo out to  ${\rm kpc}$ scales for small halo-SMBH masses, as is clearly evident from Figure~\ref{fig:r_RMS}, even for $m_{\rm ax} = 10^{-21} {\rm eV}$,
which is consistent with most current constraints on FDM 
 in dwarf galaxies. 
The effect being larger for smaller systems
is  a consequence of the fact that the 
ratio of SMBH  to halo mass is smaller for smaller systems,
the holes being lighter relative to the haloes at smaller masses. 
This combines with another effect; FDM fluctuations are 
larger in smaller systems, due to larger de Broglie wavelengths.
(Equation~\ref{eq:MBMH}). 

This  may have important consequences for observations of off-centre black holes in dwarf galaxies
(\citealp{Menez_off_cent14, Menez_off_cent16, Reine_wandering_20, Shen_2019}). 
Explanations for this phenomenon usually involve
dynamical dislocation as a result of a major merger, 
(e.g., \citealp{Comerf_BHM14, Belo_Sig_BHinD19}),  
or ejection {\it via} large gravitational 
wave recoils (e.g, \citealp{Komossa_Recoil12}) during the final stages of such mergers. 
However major mergers of dwarfs are rare, especially at lower redshifts. 
A recent 
alternative invokes sinking of smaller substructures to the 
central parent halo region, dislocating the BH in the process
(\cite{Silk_Sink20}). 

The present novel mechanism, 
involving FDM haloes, is different from the aforementioned 
attempts at explanation, as it involves
a fundamental property of FDM. It also naturally 
explains why dislocated SMBH should be more common
in small galaxies, for two  crucial reasons: 
in addition to the aforementioned SMBH-Halo mass scaling relation
favouring a large effect in smaller systems, 
the present mechanism is more efficient in dark matter-dominated 
galaxies, as we now discuss. 

Conclusions regarding SMBH expulsion from the central regions of
dark matter-dominated dwarfs are hardly affected if a modest 
baryonic component is added. Indeed, as we have seen,  assuming a baryon fraction $\sim 0.2$
has little effect. The situation is however different 
if baryons dominate the mass distribution; in this case 
dynamical friction coupling with the SMBH can 
lead the latter to sink to the centre, the FDM 
fluctuations notwithstanding.
In Section~\ref{sec:baryons} we estimated that 
limits on the axion mass are decreased 
by a factor $(\rho_{\rm halo}/\rho)^{2/3}$, 
when a baryonic component of density $\rho_* = \rho - \rho_{\rm halo}$ 
is introduced while maintaining the equilibrium, thus decreasing 
the equilibrium RMS radius by a factor $\rho_{\rm halo}/\rho$. 
Such modifications are not very large, unless $\rho_* \gg rho_{\rm halo}$.  
However the out of equilibrium inspiraling, discussed 
in previous section is significantly enhanced
already for $\rho_*$ of order $\rho_{\rm halo}$; 
as figure~\ref{fig:Reduction1} suggests, the merger rate
is much less efficiently suppressed in this case.

In Section~\ref{sec:expulsion} 
we derived the limit  
$m_{\rm ax} \ga 10^{-21} {\rm eV}$, by requiring the 
RMS displacement of the SMBH from the centre of the halo 
to be kept above  $100 {\rm pc}$ in massive systems. 
Observations reveal SMBHs at much smaller distances from the centre
than $100 {\rm pc}$. 
The significance of the $100 {\rm pc}$ scale also comes from 
the fact 
that it is about an order of magnitude larger than the combined 
sphere of influence of SMBHs of the relevant masses. 
Current simulations show that, starting at separations
a few times larger than the latter scale, a combination 
of few-body scattering and overlap of the nuclear clusters 
surrounding the black holes result in markedly accelerated 
decrease in separation,  
leading to a bound binary system (\citealp{KhanSwift16, KhanDyn18, OgiyaAcc20}). 
In this context, it becomes apparent that
if one keeps the separation above $100 {\rm pc}$, 
hierarchical SMBH growth can be effectively halted by FDM fluctuations
when  $m_{\rm ax} \ga 10^{-21} {\rm eV}$, if 
the effect of baryons is small. 
If a baryonic bulge component completely 
dominates the central dynamics, 
on the other hand,  
the  weaker constraint $m_{\rm ax} \ga 10^{-22} {\rm eV}$,
associated with few ${\rm kpc}$ scale RMS displacements, may be relevant. 
At smaller radii, 
dynamical friction coupling with the bulge stars 
can still cause the SMBH to sink. We now briefly 
discuss the robustness of this weaker constraint. 

The decomposition 
of disk galaxies into bulge, disk and halo components is complex,
and can lead to radically different results depending on assumptions, 
particularly regarding the mass-to-light ratios (e.g.,
\citealp{SOfueRot16, Bardecom_Richards18}).  
It is however unlikely that a three-dimensional 
baryonic component dominates most disk galaxies 
at scales larger than $\sim {\rm  kpc}$. 
SMBHs of masses $10^6 M_\odot \la M_B \la 10^7 M_\odot$ 
are also observed in low surface brightness galaxies (\citealp{Subra_SMBH_LSB16}).
Furthermore, in the FDM scenario, the same fluctuations 
leading to SMBH Brownian motions may also dilute the 
central density of the stellar component,
inducing a core  (\citealp{Mocz2019}). 
An axion mass of order $10^{-22} {\rm eV}$ should  therefore be firmly ruled
out. 

In general, notwithstanding possible eventual baryon domination in the central region, SMBH growth may 
be severely impeded in FDM cosmologies at higher redshifts, 
as baryons are still settling into the halo potential wells.
The impedance is likely to be far more efficient than in interacting dark matter
models, where the formation of  halo cores is sufficient to impede SMBH growth 
at higher $z$ (\citealp{Aka_SIDM20}). As we show below, even with significant contributions from the baryons, the SMBH merging process may be dramatically slowed down in FDM haloes. 
 
 In addition to the role of baryons, another
 potential uncertainty involves soliton core dynamics. 
 As already noted,  our model in not strictly valid inside the soliton core. 
 Core oscillations exist, and they can in principle 
 contribute to the stochastic dynamics 
 (\cite{Veltmaat_2018, Marsh2018}), 
 They are however likely to have different 
 characteristics than those assumed here 
 (see~\citealp{EZFCH} for discussion). 
 Nevertheless,  for the parameters of interest, 
 equations~(\ref{eq:estsmallx}) and (\ref{eq:r_sol}) show that 
 the ratio $r_{\rm sol}/r_{\rm RMS} \ll 1$; for example, at 
 $M_B = 10^7 M_\odot$, this is 
 about $1/16$ and $1/5$, for $m_{\rm ax} =10^{-22} {\rm eV}$ and
 $m_{\rm ax} =10^{-21} {\rm eV}$ 
 respectively (and varies very weakly with SMBH mass). 
 The volume enclosed by the soliton core relative to 
 that comprised by the expected RMS radius is therefore of order a   
 few times $10^{-4}$ to $10^{-2}$.  
 Furthermore, as long as the relaxation time 
 is longer than the crossing time, the motion is 
 to a first approximation ballistic over such a timescale; an SMBH 
 wandering into the core is therefore likely to enter not as  a
 result of a diffusive process, involving small energy decrements, 
 but with enough energy to exit the core on a crossing time 
 rather than be trapped
 (the situation whereby motion affected by fluctuations
 remains approximately ballistic on a dynamical timescale
 is akin to described by the orbit averaged Fokker-Planck equation~\citep{BT}). 
Finally, for a range of the relevant parameter space considered here
the SMBH may in fact grow to swallow and subsume the central soliton within
a Hubble time, in which case our calculations would be valid throughout the 
central halo (according to equations 59-60 of~\citealp{Hui_etal2017},  this should be the case
for larger  axion and halo masses).  

 Our calculations therefore suggest that an SMBH,  
 dislocated well outside the  core
 during a merger, may maintain a large  RMS displacement, 
 provided the FDM fluctuations are large enough to overcome any dynamical 
 friction coupling with baryons. 
 Given the halo density profile and the
 SMBH mass-halo mass empirical scaling relation,  
 the strength of the fluctuations depends only on the axion mass. 
 In terms of astrophysical 
 consequences, the baryon distribution is therefore a more crucial
 issue than the question of proper theoretical 
 modelling of core oscillations.

\section{Conclusion}
\label{sec:conclusions}

We have considered a central supermassive black hole immersed in a fuzzy dark matter halo of ultra-light axions. We argue that the SMBH will undergo Brownian motion induced by the FDM fluctuations due to the large de Broglie wavelength of axions. In effect, the FDM halo acts as a heat bath that expels the SMBH  from the center of the halo.
Our analysis applies outside the soliton core of FDM haloes. {\rm The physical 
picture therefore 
envisions a black hole that is initially displaced from the galaxy centres during a merger. 
We then examine whether its subsequent stochastic motion is characterized by 
a large RMS displacement.} 

The evaluation of RMS displacements assumes an SMBH in thermal equilibrium 
with the FDM heat bath. It thus presumes a small enough relaxation
time, permitting equilibration. 
The relaxation timescale of the SMBH-FDM system is depicted in Figure \ref{fig:tau}. It is less than a Hubble time for both Sagittarius $A^*$ and  heavier SMBHs for all axion masses considered here. SMBHs in dwarfs are probed by our model 
for axion masses in the range $m_{\rm ax} \gtrsim 10^{-21}{\rm eV}$. 
FDM-induced SMBH ejections for such 
systems still occur for smaller axion masses, 
but the relaxation time-scales are larger than a Hubble time.  In these cases, the mean RMS displacements calculated here, assuming equilibration of the SMBH-FDM system, are overestimates that predict excursions larger than the expected virial radii of the host haloes.
Out-of-equilibrium analysis is required to probe such cases with relaxation time-scales larger than a Hubble time.

In general,  SMBH displacement from the centre is found to be much larger for small galaxies (equation (\ref{eq:r_RMS}) and Figure \ref{fig:r_RMS}). This is a consequence of the scaling relation between SMBH and halo mass, with the central black holes predicted to be relatively lighter for less massive haloes
(\ref{eq:MBMH}). This combines with the fact that  FDM fluctuations are larger in smaller systems, due to larger de Broglie wavelengths.
FDM expulsion of SMBHs could thus explain the low detection rate of AGN in dwarf galaxies \citep{Lupi_2020}, as the accretion would be mostly suppressed for an off-set black hole. Such off-set black holes are not excluded by observations even at low redshift (e.g., \citealp{Shen_2019}). 
For dark matter-dominated dwarf galaxies, this conclusion is relatively robust
and involves a 
fundamental physical phenomenon: 
fluctuations associated with the large de Broglie wavelengths of  ultralight
axions. In luminous  galaxies, the role of the baryon distribution becomes 
important.

Our analysis further suggests that a lower bound on axion masses can be placed by heavier SMBHs, like Sagittarius $A^*$. { For the SMBH  to be kept at $100 {\rm pc}$ from the centre requires an axion mass $m_{\rm ax} = \ga 10^{-21} {\rm eV}$, while displacement of 
$1 {\rm kpc}$ leads to $m_{\rm ax} \ga 2~10^{-22} {\rm eV}$. 
In addition to 
well-centered 
SMBHs being observed at smaller radii, the significance of the $100 {\rm pc}$
scale pertains in particular to the effective stalling of mergers. 
Numerical simulations show that when the separation of SMBH pairs
in merging systems becomes significantly smaller, a markedly accelerated 
approach takes place,  leading to bound binaries as the radii of influence 
overlap (e.g., \citealp{KhanDyn18}).  
A boson mass guaranteeing such separation may  effectively
impede any SMBH growth in a hierarchical structure formation scenario, provided 
the dark matter is dominant in the central region.

The weaker constraint $m_{\rm ax} \ga 10^{-22} {\rm eV}$, deduced from an assumed 
RMS displacement  of a few ${\rm kpc}$, 
on the other hand,  corresponds to a situation 
whereby a bulge component dominates the dynamics below this scale. In such a  case, dynamical coupling between SMBHs and the stellar distribution may still lead to
an inspiraling SMBH inside the bulge (Section~\ref{sec:inter}). 
This may have interesting consequences, in this novel context, 
for the well-probed correlation 
between bulge and SMBH mass. 

In general, however, even for significant baryon mass fractions,} 
SMBH-binaries formed in galaxy mergers will tend to be softer due to the FDM fluctuations. The merging of such binaries will be significantly impeded; the event rate reduction factor depicted in Figure \ref{fig:Reduction} 
predicts that, for a universal baryon fraction, 
the LISA event rate of SMBH mergers will be reduced by at least an order 
of magnitude within FDM haloes. The impeding effects of SMBH mergers apply for the whole halo-SMBH and axion mass range, but 
are more pronounced for lower mass SMBHs and lighter axion masses.

{ The suppression of SMBH growth 
in FDM models is expected to be generally larger at higher 
redshifts, as the baryons are still settling into the potential wells of haloes. 
This effect may also be expected to be far stronger than that recently reported 
in the context of self-interacting dark matter cosmologies (\citealp{Aka_SIDM20}). 
In this case,  the suppression 
is simply a result of the formation of 
cored dark matter haloes, decreasing the dynamical friction coupling 
with the dark matter, especially at higher redshift. 
Whereas in the FDM scenario,  the suppression would 
equilibrium corresponding to large displacements from the centre.}  

Further work on the dynamics of SMBH-FDM systems, 
and its astrophysical implications, could include 
models allowing for more precise quantification and parameter variation. 
This may require moving beyond the current simple nearly isothermal fit 
and  quasi-particle approximation, appropriate for 
FDM systems with Maxwellian distributions. 
A more general formulation  could involve numerical simulations or 
Monte Carlo models of the Brownian motions. 
More detailed dynamical modelling may
incorporate core oscillations (e.g.~\cite{Veltmaat_2018}), to include the effect of fluctuations near and inside the soliton core, where the model of~\cite{EZFCH}
employed here is not strictly valid. 
From an astrophysical point of view, more ambitious
extensions would include a detailed examination of the competition 
and interplay between baryon and FDM components in determining the dynamics 
of SMBH.



\section*{Acknowledgements}
We thank Jens Niemeyer and Adi Nusser for feedback on the manuscript. 
This project was supported financially by the Science
and Technology Development Fund (STDF), Egypt. Grant
No. 25859.




\bibliographystyle{mnras}
\bibliography{FDM_halo_BH_Exp-Imp} 




\appendix


\section{Core radius}
\label{app:corad}

\begin{figure}
 	\includegraphics[width=0.49 \textwidth]{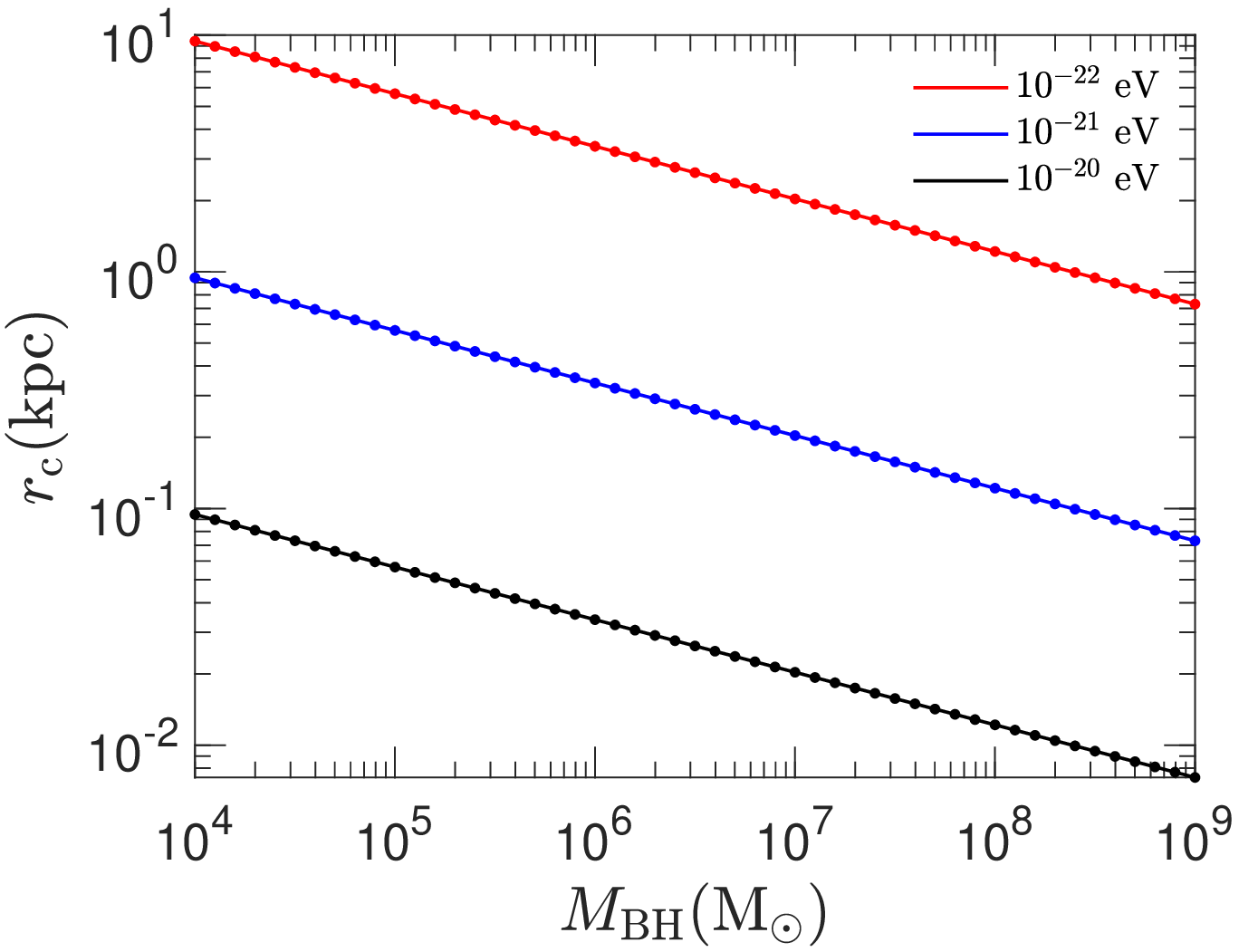}
 	\caption{The characteristic halo radius $r_c$ with respect to the SMBH mass for three different axion masses. The dots correspond to the exact Eq. (\ref{eq:r_c_gen}) and the solid line to the fit (\ref{eq:r_c_fit}).}
 	\label{fig:r_c}
 \end{figure}

We wish to determine the value of the softening radius $r_c$ which matches the core and halo profiles at some distance $r_\star = \alpha r_\text{sol}$ for some predetermined $\alpha$. We get 
\begin{equation}\label{eq:r_c_gen}
r_c = r_\text{sol}
\sqrt{\frac{1}{2}\frac{\rho_\text{I}}{\rho_\text{S}}
	\left(1-2\frac{\rho_\text{S}}{\rho_\text{I}}\alpha^2 + 
	\sqrt{1-\frac{8}{3}\frac{\rho_\text{S}}{\rho_\text{I}}\alpha^2} \right)
}
\end{equation}
where the two characteristic densities are 
\begin{equation}
\rho_\text{S} = \rho_{\rm core}(\alpha r_\text{sol}),
\quad\rho_\text{I} = \frac{3\sigma_H^2}{2\pi G r_\text{sol}^2}.
\end{equation}
The soliton profile (\ref{eq:rhosol}) is valid approximately up to $3r_\text{sol}$ \citep{Mocz2019} and therefore we set $\alpha = 3$ for the matching with the isothermal fit(\ref{eq:rho_halo}). 
Assuming relations (\ref{eq:HMSIG}), (\ref{eq:MBMH}) between BH mass, halo mass and velocity dispersion, the equation (\ref{eq:r_c_gen}) is accurately fit for $\alpha=3$ by 
\begin{equation}
r_c = 2.03~{\rm kpc} \left(\frac{m_{\rm axion}}{10^{-22}{\rm eV}}\right)^{-1} \left(\frac{M_{\rm BH}}{10^7M_\odot}\right)^{-2/9},
\label{eq:r_c_fit}
\end{equation}
shown in Figure \ref{fig:r_c}.

\section{Velocity dispersion profile}
\label{app:velprof}

We wish to evaluate the vecloity dispersion profile associated 
with the density distribution~(\ref{eq:rho_halo}) and potential~(\ref{eq:Phi}). 
In general, for spherical systems with isotropic velocities the Jeans equation
\begin{equation}
\frac{d (\rho \sigma^2)}{d r} = - \rho \frac{d \Phi}{d r}
\end{equation}
applies.  This has solution
\begin{equation}
\rho (r) \sigma^2 (r) = - \int  \rho (r) d \Phi +  C. 
\end{equation}
For the density profile~(\ref{eq:rho_halo})
$\rho (r)^{-1} \sim r^2$ at large radii, and 
therefore $C$ must be zero if the velocity dispersion is not to diverge at such radii. 
By evaluating the integral for the density and potential distributions given by 
~(\ref{eq:rho_halo}) and ~(\ref{eq:Phi}) one then finds 
\begin{equation}
\sigma (r)^2 = \sigma_H^2 \left[\frac{x^2 + 2}{x^2 + 3}\right],
\end{equation}
The velocity dispersion thus tends to that of the singular isothermal sphere  $\sigma_H$ 
at large radii and differs from this by at most a factor of $\sqrt{2/3}$ or about $18 \%$.

The logarithmic derivative of the density distribution~\ref{eq:rho_halo} is given by
\begin{equation}
\frac{d \ln \rho}{d \ln x} = - 2 ~\frac{x^4 + 5 x^2}{x^4 + 4 x^2 + 3}.
\end{equation}
Thus our density profile is flatter than the NFW cusp for radii $r \la 0.6 r_c$ and 
has similar slope for larger radii of relevance to our calculations (up to a few $r_c$). 
 
\section{Coulomb logarithm} 
\label{app:Coul}
We estimate the argument of the Coulomb logarithm
entering equations~(\ref{eq:disp_inc}) and (\ref{eq:tau}). In the context of 
the stochastic model applied here,  this is the ratio of maximal 
and minimal fluctuation scales of a Gaussian random 
field. It is thus given in terms of the maximal and minimal
wavelength modes as
$\Lambda = \lambda_{\rm max}/ \lambda_{\rm min}$. 
Following Appendix~E of \cite{EZFCH}, we set the minimal scale 
to correspond to the size associated with the effective 
mass of the FDM granules: $m_{\rm eff} = \frac{4}{3} \rho_{\rm halo} \lambda_{\rm min}^3$. 
Using~(\ref{eq:meff}) gives
\begin{equation}
\lambda_{\rm min} = \left(\frac{3}{4}\right)^{1/3} \pi^{1/6} \frac{\hbar}{m_{\rm ax} \sigma_{\rm halo}}.
\label{eq:lambdamin}
\end{equation}
In \cite{EZFCH}, the main concern was the effect of FDM fluctuations on disks. 
A cutoff was thus introduced to exclude longer wavelengths, as these would affect the disk nonlocally, 
possibly leading to the excitation of  global modes, with consequences 
not captured by the stochastic model. In the present situation of an SMBH orbiting 
in the central region, all modes supported by the halo heat bath 
should contribute. 
We therefore set largest scales to correspond to the virial radius of the halo. 
For $\lambda$CDM cosmologies this may be approximated at redshift $z= 0$ by
(e.g., Appendix of \citealp{Navarro1997})
\begin{equation}
\lambda_{\rm max} = R_H \approx   200~{\rm kpc}  \left( \frac{M_H}{10^{12} M_\odot}\right)^{1/3}.
\label{eq: virialr}
\end{equation}
Using equation~(\ref{eq:lambdamin}) in conjunction with (\ref{eq:HMSIG}), 
and neglecting the modest departures of $\sigma_{\rm halo}$ from $\sigma_H$ 
(cf. equation \ref{eq:sigma_halo}), one  gets
\begin{equation}\label{eq:Lambda}
\Lambda \approx  1000~\left(\frac{m_{\rm ax}}{10^{-22}}\right) \left(\frac{M_H}{10^{12} M_\odot} \right)^{2/3}. 
\end{equation}
 
 \section{The advent of effective quasiparticle picture}
\label{app: effective}

The description of FDM fluctuations used here involved 
an effective mass, velocity dispersion and Coulomb 
logarithm. However it is important to point out that this effective 
picture does not
require the fluctuations 
to behave as long lived classical particles. 
Indeed, the standard Chandrasekhar estimates of relaxation and dynamical friction are quite conveniently recovered by formulating 
the problem in Fourier space. In this way 
it is possible to describe classical two body relaxation in terms 
of a superposition of 
waves, even though, physically speaking, one is describing a 
medium of discrete  particles  ((\citealp{EZFCH, BOFT}).
In the same generic manner one can also describe the effect 
of fluctuations arising from fully 
developed turbulence, even though one deals with a fluid
(\citealp{EZFC}). 

In this context, \cite{EZFCH} applied the model initially developed  
for turbulent media to systems of classical particles and to interfering 
FDM waves.  They thus found it  possible  to recover the results of  
classical two body relaxation, in addition to an analogous 
effect due to the interfering FDM de Broglie waves. 
The effective description of FDM fluctuations transpired by relating the results 
of the latter analysis to 
those of classical two body relaxation. But the quasiparticles, that thus arise, are 
no more classical particles than the Fourier modes entering into the 
derivation of classical two body relaxation are 'real' (physical) classical waves.
As we now discuss, they simply correspond to a cutoff scale of a power spectrum 
of interfering waves, which in fact has a white noise form on larger scales.  

\subsection{Density fluctuations}

We start  with 
the density contrast power spectrum. Following~\cite{EZFCH}, this 
is given by
\begin{equation}
\mathcal{P} ({\bf k}, t)\! = \!\frac{(2 \pi)^3}{\rho_0^2} \!\!\int\!\! d {\bf v_1} d {\bf v_2} f (\! {\bf v}_1\!) f (\!{\bf v}_2\!)
\delta_D ({\bf k} - m_\hbar\! {\bf v}_d) e\!^{-i  m_\hbar ({\bf v}_c . {\bf v}_d) t},
\label{eq:PS_Ax}
\end{equation}
where 
\begin{equation}
m_\hbar = 2 m_{\rm ax}/\hbar,
\label{eq:mhbar}
\end{equation}
${\bf v}_i = \hbar {\bf k}_i/ m$ are de Broglie wave packet group velocities, 
and ${\bf v}_c$ and ${\bf v}_d$ correspond to the sum and differences of the phase velocities
of interfering waves; such that 
$2 {\bf v}_c = {\bf v}_1 + {\bf v}_2$, 
and $2 {\bf v}_d = {\bf v}_1 - {\bf v}_2$.  For a Maxwellian velocity distribution 
the equal time power spectrum was shown to agree nicely with power spectra inferred 
from numerical simulation; the spectra in 
both the theory and simulations simply correspond to a white  noise power spectrum 
with cutoff scale characteristic of the typical de Broglie wavelength. Indeed, for 
a system with velocity dispersion $\sigma$, the RMS fluctuations on scale $R$ 
were found to decrease as 
\begin{equation}
\sigma_R^2 = \left( \frac{2}{(R \sigma m_\hbar)^2 + 2} \right)^{3/2},
\end{equation}
characteristic of white noise decline on scales 
$R \gg \sigma m_\hbar$. The density 
correlation functions (especially in the temporal regime) 
also agreed with published simulation results.

\subsection{Dynamical effect in terms of quasiparticles}

The quasiparticle picture arises from the aforementioned power spectrum cutoff 
as follows. The density fluctuations give rise to force fluctuations (through the Poisson equation). 
When assumed to form a random Gaussian field, these can be quantified through 
the force correlation function. \cite{EZFCH} find this to be given by
\begin{align} 
\nonumber
\langle {\bf F} (0, 0) .  {\bf F} (r, & t)\rangle = \\
& \left(\frac{4 \pi G}{m_\hbar}\right)^{2}
\!\!\! \int \! \frac{e^{i m_\hbar ({\bf v}_d . {\bf r}  - {\bf v}_c . {\bf v}_d t)}}{v_d^2}
f ({\bf v}_1) f ({\bf v}_2) d {\bf v}_1 d {\bf v}_2. 
\label{eq:fullqcorr}
\end{align}
Assuming isotropy, the velocity dispersion arising from the fluctuations 
after time $T$, for a classical test particle moving with velocity ${\bf v}_p$
---  $\langle (\Delta v_p)^2 \rangle = 2 \int_0^T (T - t)  \langle {\bf F}(0) . {\bf F}(t) \rangle d t$  --- can then be written as
\begin{align}
\nonumber
\langle (\Delta v_p)^2 \rangle = A
\!\int \!\! & (T-t) f ({\bf v}_c + {\bf v}_d)   f ({\bf v}_c - {\bf v}_d)\\ 
& \times
\frac{\sin \left(m_\hbar v_d |{\bf v}_p - {\bf v}_c| t\right)}{m_\hbar v_d |{\bf v}_p - {\bf v}_c| t}~
d {\bf v}_c d v_d d t,  
\label{eq:axforccorr}
\end{align}
where $A =  2 (8 \pi)^3 G^2 m_\hbar^{-2}$.

In~\cite{EZFCH} it is argued that, as 
the integral is dominated by the region
$v_d \ll  v_c$,  both distribution functions 
figuring in this equation can be approximated  
as $f ({\bf v}_c)$. If this is the case, in the diffusion (large $T$) limit, one finds
\begin{equation}
\langle (\Delta v_p)^2 \rangle = \left(\frac{4 \pi}{m_\hbar}\right)^3 8 \pi G^2 T   \ln \Lambda 
\int d {\bf v}_c  \frac{f^2 ({\bf v}_c)}{|{\bf v}_p - {\bf v}_c|},
\label{eq:ax_disp}
\end{equation}
where here
\begin{equation}
\Lambda = \frac{v_{dx}}{v_{dm}},
\label{eq:LFDM}
\end{equation}
is a ratio of maximal and minimal speeds, which are proportional to
corresponding wavenumbers by virtue of the de Broglie relation. 

Introducing 
\begin{equation}
m_{\rm eff} = 
\left(\frac{4 \pi}{m_\hbar}\right)^3
\frac{\int f^2 ({\bf v}) d {\bf v}}{\int f ({\bf v}) d {\bf v}},  
\label{eq:meffO}
\end{equation}
where $m_\hbar$ is given by (\ref{eq:mhbar}) and
\begin{equation}
f_{\rm eff} ({\bf v})  =  \frac{\int f ({\bf v}) d {\bf v}}{\int f^2 ({\bf v}) d {\bf v}}  f^2 ({\bf v}),  
\label{eq:feff}
\end{equation}
makes (\ref{eq:ax_disp})  formally equivalent 
to the expression describing the increase 
in velocity dispersion due to fluctuations
arising from a distribution of classical field 
particles of mass $m$ (\citealp{EZFCH, BT} Appendix L)
\begin{equation}
\langle (\Delta v_p)^2 \rangle = 8 \pi m G^2 T \ln \Lambda \int d {\bf v} \frac{f ({\bf v})}{|{\bf v}_p - {\bf v}|}.
\label{eq:velydist}
\end{equation}
In the standard derivation $\Lambda$ corresponds to a ratio of maximal and minimal
impact parameters, while when the fluctuating field is Fourier analysed it 
corresponds to a ratio of maximal and minimal  cutoff wavelengths
(or wavenumbers).

The approximation that leads to this formal equivalence between FDM 
and classical particle systems, effectively  
assumes a long wavelength limit in the power spectrum of fluctuations,  which  
then simply corresponds to white noise. The small scale cutoff 
at the characteristic de Broglie wavelength  
is accounted for by the large velocity (small spatial scale) 
cutoff determined in the Coulomb logarithm (\ref{eq:LFDM}). 
This approximation would be exact for 
a distribution function in the form of a step function, with sharp cutoff
at the characteristic wavelength, corresponding to an effective 
'size' for the quasiparticles;  
for in this case the $1/v_d$ in equation (\ref{eq:axforccorr})
need not be integrated over the 
distribution functions. Its independent integration leads to the Coulomb logarithm, 

In~\cite{EZFCH}, it is furthermore argued that this is also a good approximation for Maxwellians, which are 
roughly constant for smaller speeds, before
quickly cutting off beyond a certain characteristic value (corresponding to 
larger wavenumbers and smaller spatial scales). 
We now show explicitly that this is indeed the case.  

\subsection{Validity for Maxwellian distributions}

For the Maxwellian distribution 
\begin{equation}
f (v) = \frac{\rho_0}{(2 \pi \sigma^2)^{3/2}} e^{-\frac{v^2}{2 \sigma^2}}, 
\label{eq:Max}
\end{equation}
equation~(\ref{eq:axforccorr})
becomes
\begin{align}
\nonumber
\langle (\Delta v_p)^2 \rangle = B~
\!\int \!\! & (T-t) e^{- \frac{1}{\sigma^2}[({\bf v}_c + {\bf v}_d)^2 + ({\bf v}_c - {\bf v}_d)^2]}\\
& \times \frac{\sin \left(m_\hbar v_d |{\bf v}_p - {\bf v}_c| t\right)}{m_\hbar v_d |{\bf v}_p - {\bf v}_c| t}~ 
d {\bf v}_c d v_d d t,  
\label{eq:axforccorrM}
\end{align}
where $B = A~\rho_0^2/(2 \pi \sigma^2)^{3}$.  

Expanding the squares in the exponential, then  
 integrating over time and taking the diffusion limit
 ($T \gg  m_\hbar v_{d} |{\bf v}_p - {\bf v}_c|$), 
 this becomes
\begin{equation}
\langle (\Delta v_p)^2 \rangle = \left(\frac{4 \pi}{m_\hbar}\right)^3 8 \pi G^2 T   \int \frac{e^{-\frac{v_d^2}{\sigma^2}}}{v_d}  d v_d 
\int d {\bf v}_c  \frac{f^2 ({\bf v}_c)}{|{\bf v}_p - {\bf v}_c|}.
\label{eq:ax_DMax}
\end{equation}
This is the same as equation~(\ref{eq:ax_disp}) 
above, except that the Coulomb logarithm is replaced by integration of $1/v_d$ weighed by 
the exponential (this result was arrived at from a different procedure in~\citealp{BOFT}). 

To compare~(\ref{eq:ax_DMax}) with (\ref{eq:ax_disp}), we explicitly evaluate 
the integral of $1/v_d$.
As the contribution from larger $v_d$ 
is now naturally cutoff by this weighing, we need not impose an 
artificial high speed (wavenumber) cutoff;  
thus we may set $v_{dx} \rightarrow \infty$. The minimal cutoff 
is still limited by the system size (as described in the previous Appendix). 
We can thus evaluate this integral as follows:
\begin{equation}
\int_{v_{dm}}^{\infty}  \frac{e^{-\frac{v_d^2}{\sigma^2}}}{v_d}  d v_d  = 
-\frac{1}{2} {\rm Ei} \left(- \frac{v_{dm}^2}{\sigma^2}\right) 
\simeq \ln \left(\frac{\sigma}{v_{dm}}\right) - 0.29,  
\end{equation}
where ${\rm Ei}$ is the exponetial  integral, 
0.29 stands for an approximation to half the Euler–Mascheroni constant, and where we ignore terms of order $(v_{dm}/\sigma)^2$
and higher; as from (\ref{eq:lambdamin}),
$\sigma \simeq \sigma_{\rm halo}$  approximately corresponds to the minimal 
de Broglie scale, while 
$v_{dx}$ is associated with the much larger maximum scale
(given by~\ref{eq: virialr}). 

Setting
\begin{equation}
    v_{dx} = \frac{\hbar}{m_{\rm ax} \lambda_{\rm max}},  
\end{equation}
with $\lambda_{\rm max}$ given by (\ref{eq: virialr}),
gives a ratio 
\begin{equation}
\frac{\int_{v_{dm}}^{\infty} e^{-\frac{v_d^2}{\sigma^2}} d v_d}{\ln \Lambda}
= \frac{\ln  \left(\frac{3}{4}\right)^{1/3} \pi^{1/6} \Lambda -0.29}{\ln \Lambda}
=
1  - \frac{0.19}{\ln \Lambda}.
\end{equation}
With $\Lambda$ given by (\ref{eq:Lambda}), the second term
in the last equality is small. Which means that the exact 
integral over the distribution function 
quite accurately corresponds in this case with 
the approximation in terms of Coulomb logarithm, from which 
the correspondence with classical relaxation, and the notions of 
effective mass and distribution function, arose. 

Thus, for a Maxwellian distribution, the quasi-particle 
picture is accurate, with the exact results recovered 
{\it via} a judicious choice of the Coulomb logarithm, 
with a natural spatial cutoff scale corresponding to
the characteristic 'size' of effective  quasiparticles. 
This cutoff is directly related to that in the power spectrum of 
density fluctuations, which matches well the results of numerical 
simulations.


\bsp	
\label{lastpage}
\end{document}